# Spatially modulated 'Mottness' in $La_{2-x}Ba_xCuO_4$

P. Abbamonte[1,a], A. Rusydi[1,3,b], S. Smadici[1], G. D. Gu[1], G. A. Sawatzky[3,4], & D. L. Feng[5]

[1]*National Synchrotron Light Source and Physics Department, Brookhaven National Laboratory, Upton, New York 11973, USA*

[3]*Materials Science Centre, University of Groningen, 9747 AG Groningen, The Netherlands*

[4]*Department of Physics and Astronomy, University of British Columbia, Vancouver, British Columbia V6T-1Z1, Canada*

[5]*Department of Physics, Fudan University, Shangai 200433, China*

**Stripe phases were predicted[1,2,3,4] to arise in doped antiferromagnets through competition between magnetism and the kinetic energy of mobile carriers (typically holes). In copper-oxides the main experimental evidence for stripes is neutron scattering from $La_{1.48}Nd_{0.4}Sr_{0.12}CuO_4$[5] (LNSCO) and $La_{1.875}Ba_{0.125}CuO_4$[6] (LBCO) which reveals coexisting static spin and charge order whose wavelengths differ by a factor of two, reminiscent of charged rivers separating regions of oppositely-phased antiferromagnetism. A neutron is an electrically neutral object, however, so does not detect charge but rather its associated lattice distortion[7]; it is not known if the "stripe" phase in LNSCO and LBCO actually involves ordering of the doped holes. Here we present a study of the charge order in LBCO with resonant soft x-ray scattering (RSXS). We observe giant resonances at both the mobile carrier and upper-Hubbard band features in the O$K$ edge[8]. These demonstrate a substantial modulation in the doped hole density as well as the amount of spectral weight near the correlated gap[8, 9], i.e. the degree of "Mottness"[10]. The peak-to-trough amplitude of the valence modulation is estimated to be 0.063 holes, which if interpreted with a model of the stripe form factor[11] suggests an integrated area of 0.59 holes under a single stripe. While only an estimate, this number agrees with what is expected for half-filled stripes.**

The charge/spin superstructure in LNSCO[5] and LBCO[6] appears only in the low temperature tetragonal (LTT) phase, is most stable at $x = 1/8$ and coincides with an anomalous suppression of $T_c$ [12]. This phase is frequently interpreted as (quasi) static stripes which have been pinned by the LTT distortion. The charge reflections observed with neutron scattering are weak (~6 times less intense than the magnetic reflections) since neutrons detect only the lattice distortion which was estimated to be only ~ 0.004 Å [7]. However one assumes the hole modulation itself is significant. We point out, however, that the spin density wave in elemental Cr also exhibits half-wavelength charge reflections which are weaker by a factor of ~4.1 and represent a distortion of similar size[13]. So in the neutron Bragg peaks alone there is no clear difference between the phenomenon in LNSCO and a simple spin density wave. To determine if the doped holes are actually involved we have studied LBCO with resonant soft x-ray scattering near the O$K$ ($1s \rightarrow 2p$) and Cu$L_{3/2}$ ($2p_{3/2} \rightarrow 3d_{x2-y2}$) edges, which provide direct sensitivity to valence electron ordering[14,15,16,17,18,19,20].

Single crystals of La$_{2-x}$Ba$_x$CuO$_4$ with $x = 1/8$ were grown by the floating zone method[21]. The sample used in this study had $T_c = 2.5$ K indicating suppressed superconductivity and stabilized spin/charge order. The sample was cleaved in air revealing a surface with (0,0,1) orientation. RSXS measurements were performed on beam line X1B at the National Synchrotron Light Source using a 10-axis, ultrahigh vacuum-compatible diffractometer. The sample was cooled with a He flow cryostat connected via Cu braids, providing a base temperature of 18 K. X-ray absorption spectra (XAS) were measured *in situ* in fluorescence yield mode at the O$K$ and Cu$L_{3/2}$ edges and found to be consistent with previous studies[8] (see Figs. 1, 3$a$). We will denote reciprocal space with Miller indices ($H,K,L$), which represent a momentum transfer $\mathbf{Q} = (2\pi/a\ H, 2\pi/b\ K, 2\pi/c\ L)$ where $a = b = 3.788$ Å, $c = 13.23$ Å. The incident x-ray polarization depends on $\mathbf{Q}$ but was approximately 60° from the Cu-O bond for measurements at both edges.

The O$K$ XAS in the cuprates exhibits a mobile carrier peak (MCP) at 528.6 eV, corresponding to transitions into the doped hole levels, and an upper Hubbard band (UHB) peak at 530.4 eV, corresponding to transitions into the $d$ level of a neighboring Cu (see Figs 1 & 3). As the hole density is increased spectral weight is transferred from the UHB to the MCP as the number of sites opposing electron addition is decreased[8,9,22]. In this letter we will take the spectral weight in the MCP to be a measure of the local hole density and that in the UHB to be a measure of the "Mottness"[10], i.e. the degree to which there is a functioning Hubbard $U$. The Cu$L_{3/2}$ edge exhibits a main peak at 933 eV and ligand hole side band at 934.3 eV which grows with doping. We will take the sideband to be a measure of the hole density, similar to the MCP, and the main peak to be a measure of distortions in the Cu sublattice.

Static charge "stripe" correlations were initially detected at the Cu$L_{3/2}$ resonance and mapped in the ($H$,0,$L$) plane, shown in Fig. 2. In agreement with hard x-ray studies[23,24] the scattering was sharp along $H$ but rod-like along $L$ (correlation lengths $\xi_a \approx 127a$, $\xi_c \approx 2c$), indicating quasi-long range order in the CuO$_2$ plane but weak coupling between planes. A maximum intensity of 175 Hz on a fluorescence background of 425 Hz was observed at (0.251, 0, 3/2) [25], the half-integer $L$ indicating that charge order is offset by $2a$ between successive unit cells, presumably to minimize Coulomb repulsion. No scattering could be detected off-resonance above the fluorescence background.

In Fig. 1 we display a "resonance profile", i.e. the intensity of charge scattering as a function of energy compared to x-ray absorption spectra. For geometric convenience the O$K$ and Cu$L_{3/2}$ data were taken with $L = 0.72$ and $L = 1.47$, respectively. The charge scattering at the Cu$L_{3/2}$ edge is offset from the absorption maximum by 0.2 eV and likely arises from distortions in the Cu sublattice. The O$K$ scattering exhibits strong resonances at both the MCP and UHB features, indicating that the carrier density and the degree of Mottness are both modulated in real space. This shows that this phase exhibits real charge order and, perhaps more significantly, is associated with a periodic restoration of a Mott state.

The charge modulation appears to be substantial but it is important to estimate its size. The integrated intensity of a Bragg reflection is proportional to the square of its structure factor, $\rho_\mathbf{Q}^{mn}(\omega) = (1/V_{cell}) \Sigma_j f_j^{mn}(\omega) \exp(i\mathbf{Q} \cdot \mathbf{r}_j)$, where $V_{cell}$ is the unit cell volume, $f_j^{mn}(\omega)$ is the x-ray form factor of atom $j$, $\mathbf{r}_j$ its position, and $m$ and $n$ are incoming and outgoing polarization indices[26]. So to estimate the hole amplitude one must first define the x-ray form factor for a doped hole.

For this purpose we separate the oxygen form factor $f_O^{mn}(\omega,p) = f_R(\omega) + p\, f_D^{mn}(\omega)$, where $f_R$ is a "raw" component common to all oxygen atoms, i.e. $f_R \to 8$ as $\omega \to \infty$, and $f_D^{mn}(\omega)$ describes the polarization-dependent spectral changes with the doped hole density, $p$. $f_D^{mn}$ has units electrons / hole and can be interpreted as the doped hole form factor. A form factors is related to the absorptive part of the index of refraction by $Im[n^m(\omega;p)] = -r_e\lambda^2/2\pi\, V_{cell}\, Im[\,\Sigma_i f_i^{mm}(\omega,p)\,]$, where $r_e$ is the classical electron radius and $\lambda$ is the x-ray wavelength, so $f_D^{mn}$ may be determined from doping-dependent XAS. Using this relationship and data from Ref. 8 (Fig. 3a) we extracted the diagonal components $Im[f_D^{mm}(\omega)]$ by solving a system of equations at each energy and retrieving the real parts by Kramers-Kronig transform. The off-diagonal components were assumed to be small. For an $x$-oriented hole we find $|f_D^{xx}| = 82$ electrons / hole on the resonance maximum, i.e. a doped hole scatters as strongly as a Pb atom.

To relate $f_D^{mn}(\omega)$ to the scattered intensity a stripe structure factor $\rho_{stripes}^{mn}$ was constructed based on the orbital pattern from a previous three-band Hubbard calculation[11] (Fig. 3c). This pattern was stacked in alternating fashion as proposed in Ref. 7 and, accounting for the incident and scattered polarizations, the structure factor was computed analytically to be $|\varepsilon_f^* \cdot \rho_{stripes} \cdot \varepsilon_i| = A\,(0.00475\ \text{Å}^{-3})$. $A$ is a fit variable which may be interpreted as the number of holes contained in one stripe a single unit cell in length ($A = 0.5$ would be consistent with half-filled stripes).

The integrated intensity of resonant scattering was determined in the ellipsoidal approximation by tuning to the MCP and carrying out linear scans along three orthogonal directions and taking the product $I = I_{peak} \times \Delta q_x\, \Delta q_y\, \Delta q_z$, where $I_{peak}$ is the peak count rate in Hz and $\Delta q_n$ is the width along direction $n$. We measured $\Delta q_x = 0.0133\ \text{Å}^{-1}$ and $\Delta q_y = 0.0169\ \text{Å}^{-1}$. The limited **Q** range of RSXS prevented complete determination of $\Delta q_z$ so we use the value from Ref. 6 of $\Delta q_z = 0.256\ \text{Å}^{-1}$, giving $I_{stripes} = 0.101$ Hz / Å$^3$. To determine the absolute scale the same measurement was done 10 eV below the O$K$ edge on the (0,0,2) Bragg reflection from a cleaved single crystal of $Bi_2Sr_2CaCu_2O_{8+\delta}$, which has a known structure factor. By comparing the two measurements we arrived at $|\varepsilon_f^* \cdot \rho_{stripes} \cdot \varepsilon_i| = 0.00282\ \text{Å}^{-3}$ or $A = 0.59$ holes. This number is only an estimate but is close to the value 0.5 expected for half-filled stripes. The estimated occupancies for various orbitals are shown in Fig. 3c. Note that despite the large integral the peak-trough amplitude is only 0.063 holes, since the doped holes are distributed over many sites.

To search for a difference between the ordering of holes and the lattice, angular scans through the charge scattering were performed at different temperatures at both the MCP and $CuL_{3/2}$ resonances (Fig. 4). Charge correlations were visible at both edges below the LTT transition at 60K, which is higher than previously published[6] because the $x$ value for our sample is closer to 1/8 [27]. No difference in temperature dependence was detected between the two edges, suggesting the holes and lattice order simultaneously.

The reader may wonder about the relationship between this study and the negative result reported in ref. 12 for optimally-doped $La_2CuO_{4+y}$ (LCO), in which the stripe order is sometimes said to be "dynamic". It is well-known that a fluctuating density can give rise to diffuse, elastic scattering, a common example being Rayleigh scattering by the atmosphere. In Ref. 14 we showed that the resonant diffuse scattering from such fluctuations in LCO is negligible, i.e. such correlations may be valid excited states but are not thermally occupied, i.e. translational symmetry is not broken even on short time scales. By contrast the charge ordering in LBCO is static and, as evident from the large amplitude and the broad spectral width, a macroscopic fraction of the doped holes are modulated. The amplitude of 0.063 is similar to a recent estimate for Nickelate stripes[20] even though the lattice in LBCO is only weakly involved. The associated modulation of high energy spectral weight is strong evidence that this phase is driven instead by a desire of the system to restore a Mott state in some regions of the sample.

Our results do *not* establish whether this phase is truly nematic or is, for example, a "checkerboard" Wigner crystal[28]. Forthcoming measurements will address this issue. In the mean time, we point out the similarity of this phenomenon to the "hole crystal" recently observed in $Sr_{14}Cu_{24}O_{41}$ [17]. The importance of Umklapp processes for the stability of a Wigner crystal on a lattice may explain the stability charge order at the commensurate doping $x = 1/8$.

**Acknowledgements** We acknowledge T. Valla for assistance with sample cleaving and discussions with S. K. Sinha, J. M. Tranquada, C. Schüssler-Langheine, P. A. Lee, and Wei Ku. This work was supported by the U.S. Department of Energy, N.W.O. (Dutch Science Foundation), and F.O.M. (Netherlands Organization for Fundamental Research on Matter).

**Competing interests statement** The authors declare that they have no competing financial interests.

**Correspondence** and requests for materials should be addressed to P.A. (abbamonte@mrl.uiuc.edu).

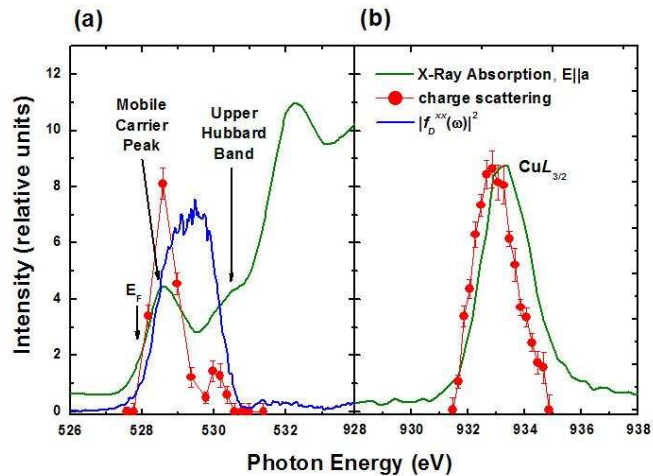

**Figure 1** A "resonance profile", i.e. the energy-dependence of the (1/4, 0, *L*) charge scattering compared to x-ray absorption spectra (XAS). **a**, Data near the O*K* edge. Green line, XAS spectrum for **E**∥**â**. Red circles, intensity of charge scattering at *L* = 0.72 showing enhancements at the mobile carrier peak (MCP) and upper Hubbard band (UHB), demonstrating a significant modulation of the doped hole density. Blue line, $|f_D^{xx}(\omega)|$ determined in Fig. 3. The energy width of $|f_D^{xx}|$ is larger than the actual data points because the photoelectron lifetime contributes in absorption but not scattering. **b,** data near the Cu*L*$_3$ edge for *L* = 1.47. A slight red shift is seen between the peak in resonant scattering compared to absorption.

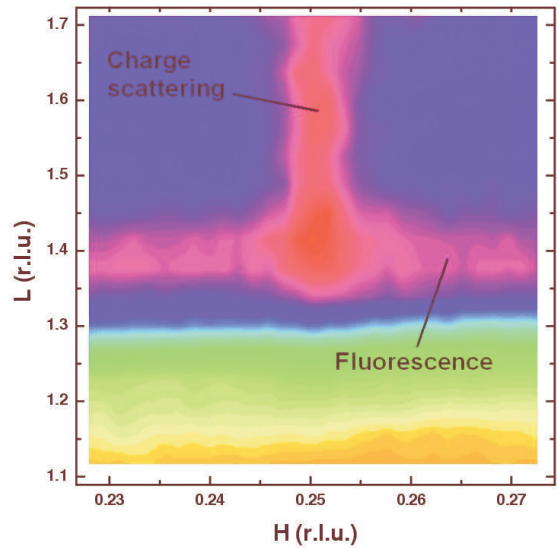

**Figure 2** Reciprocal space map of charge correlations in the (*H*,0,*L*) plane. These data were taken with the x-ray energy tuned to the Cu $L_{3/2}$ ($2p_{3/2} \to 3d_{x2-y2}$) resonance. At its maximum the charge scattering is 175 Hz on a fluorescence background of 450 Hz. As *L* is decreased the fluorescence intensity also rises as the angle of incidence on the crystal grows, reaching its maximum value at $L = 1.38$, below which the takeoff angle falls, driving the intensity to zero through sample self-absorption.

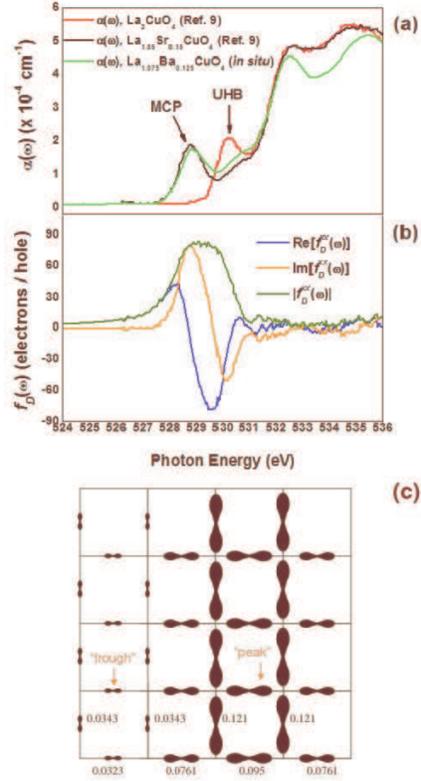

**Figure 3** Determining the x-ray form factor, $f_D^{mn}(\omega)$, for a single doped hole. **a,** x-ray absorption spectra for $La_{2-x}Sr_xCuO_4$ with $x = 0$ (red line) and $x = 0.15$ (black line) for **E**∥â, reproduced from Ref. 8. Spectra were placed on an absolute scale by fitting to tabulated values[29] far from the edge. Data from the current sample (green line) are shown for comparison. *Inset*, stripe orbital pattern from the three-band Hubbard calculation in Ref. 11 with "peak" and "trough" locations indicated. **b,** $f_D^{xx}(\omega)$ for an x-oriented hole determined by using the expression $Im[n^m(\omega;p)] = - r_e\lambda^2/2\pi\ V_{cell}\ Im[\ \Sigma_i\ f_i^{mm}(\omega,p)\ ]$ to construct a 2×2 system of equations, matrix inverting to get $Im[f_D^{xx}(\omega)]$, and Kramers-Kroning transforming. $|f_D^{xx}|$ reaches a maximum value of 82 at $\omega = 528.6$ eV, where a single hole scatters as strongly as a Pb atom.

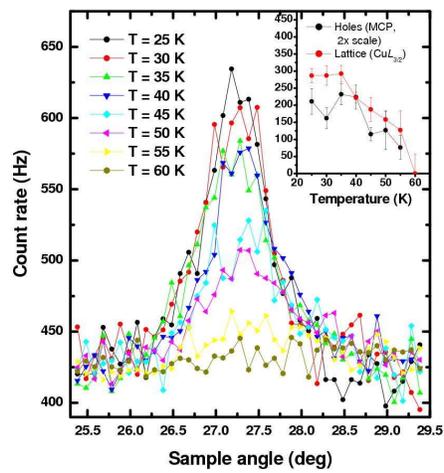

**Figure 4** Temperature dependence of the charge scattering. Charge correlations are visible below the low temperature tetragonal (LTT) phase transition temperature of 60 K. $T_{LTT}$ is higher than previously reported because *x* is closer to 1/8. Charge scattering at both edges exhibits the same temperature dependence, suggesting that the holes and lattice order together.

*a* Current address: Department of Physics, University of Illinois, Urbana, IL, 61801, USA

*b* Current address: Institute for Applied Physics, University of Hamburg, D-20355 Hamburg, Germany